\documentclass[
prd, twocolumn%
,preprintnumbers%
,showpacs ,amssymb,superscriptaddress,aps
]{revtex4}
\usepackage{graphicx}
\input{epsf}

\usepackage{amsmath,amssymb}
\usepackage{bm}

\newcommand{\dalm}{\kern1pt\vbox{\hrule height 0.9pt\hbox{\vrule width 0.9pt
\hskip 2.5pt\vbox{\vskip 5.5pt}\hskip 3pt\vrule width 0.3pt}\hrule height 0.3pt}
\kern1pt}

\begin{document}
\preprint{YITP-14-34}



\title{Observational discrimination of Eddington-inspired Born-Infeld gravity from general relativity }

\author{Hajime Sotani}
\email{sotani@yukawa.kyoto-u.ac.jp}
\affiliation{Yukawa Institute for Theoretical Physics, Kyoto University, Kyoto 606-8502, Japan
}

\date{\today}

\begin{abstract}
Direct observations of neutron stars could tell us an imprint of modified gravity. However, it is generally difficult to resolve the degeneracy due to the uncertainties in equation of state (EOS) of neutron star matter and in gravitational theories. In this paper, we are successful to find the observational possibility to distinguish Eddington-inspired Born-Infeld gravity (EiBI) from general relativity. We show that the radii of neutron stars with $0.5M_\odot$ are strongly correlated with the neutron skin thickness of ${}^{208}$Pb independently of EOS, while this correlation depends on the coupling constant in EiBI. As a result, via the direct observations of radius of neutron star with $0.5M_\odot$ and the measurements of neutron skin thickness of ${}^{208}$Pb by the terrestrial experiments, one could not only discriminate EiBI from general relativity but also estimate the coupling constant in EiBI.
\end{abstract}

%
\maketitle
\section{Introduction}
\label{sec:I}
Up to now, several modified theories of gravity are proposed, in spit of the fact that general relativity has been successful to explain the phenomena and experiments in weak-field regime such as solar system \cite{W1993}. Meanwhile, the tests of general relativity in strong-field regime are quite poor. This could be one of the reasons why the modified gravitational theories are considered. Additionally, in order to explain the unsolved issues such as singularities in cosmology and/or inside black holes, one might take into account the correction due to the higher order curvature. Anyway, since the technology is developing more and more, one will be able to observe compact objects with high accuracy and use it as tests of modified gravity \cite{P2008,SK2004,S2009,YYT2012}.

Among the several modified gravitational theories, Eddington-inspired Born-Infeld gravity (EiBI) \cite{EiBI} has recently attracted attention in the context to avoid the big bang singularity \cite{AF2012,BFS12013}, while EiBI becomes equivalent to general relativity in vacuum. EiBI is based on the Eddington action \cite{E1924} and the non-linear electrodynamics of Born and Infeld \cite{BI}, where the metric and the connection are considered as independent fields, as in Palatini-type approach to general relativity. EiBI can deviate from general relativity only when the matter exists, and one can expect the significant deviation especially in high-density region. That is, the compact objects might be suitable laboratories to probe the gravitational theory. Previously, there are several attempts to examine the structures of compact objects in EiBI \cite{PCD2011,PDC2012,SLL2012,SLL2013,HLMS2013}, which showed the significant deviations in stellar properties from the predictions in  general relativity, depending on the coupling constant. 
We remark some of the problems associated with the EiBI, i.e.,  the appearance of the curvature instabilities at the surface of polytropic stars is pointed out \cite{PS2012}, while the validity of the continuous fluid approximation adopted in astrophysical and cosmological studies is also discussed \cite{A2012}.

However, the stellar structures also depend on the unfixed equation of state (EOS), which is a relation between the pressure and density of nuclear matter. That is, it must be generally difficult to resolve the degeneracy due to the uncertainties in EOS and in gravitational theory, even if one would measure the mass and radius of neutron star.

In this paper, we find the observational possibility to distinguish EiBI from general relativity via the measurements of neutron skin thickness by the terrestrial experiments, in addition to the astronomical observations of neutron stars.
Since the neutron stars are also regarded as neutron-rich giant-nuclei, the neutron star radius could be correlated with the properties of neutron-rich nuclei. In fact, in general relativity, it is suggested that the radii of neutron stars with $0.5M_\odot$ are strongly correlated with the neutron skin thickness \cite{CH2003}.

In particular, we adopt the realistic EOSs, which are consistent with the empirical date of masses and radii of stable atomic nuclei obtained from the terrestrial experiments, and show that the radii of neutron stars with $0.5M_\odot$ can be written as a linear function of the neutron skin thickness of ${}^{208}$Pb independently of the adopted EOSs. Additionally, this linear correlation depends strongly on the coupling constant in EiBI. Therefore, one could distinguish EiBI from general relativity via the measurements of neutron skin thickness and the radii of neutron stars with $0.5M_\odot$. Furthermore, we are also successful to estimate the value of coupling constant in EiBI as a function of neutron skin thickness of ${}^{208}$Pb and stellar radius with $0.5M_\odot$. With this estimation, at least, one may be able to observationally see how valid general relativity is. In this paper, we adopt geometric units, $c=G=1$, where $c$ and $G$ denote the speed of light and the gravitational constant, respectively, and the metric signature is $(-,+,+,+)$.

\section{Eddington-inspired Born-Infeld gravity}
\label{sec:II}
EiBI proposed by Ba\~nados and Ferreira \cite{EiBI}, is described with the action
\begin{equation}
  S = \frac{1}{8\pi\kappa} \int d^4x \left(\sqrt{|g_{\mu\nu} + \kappa R_{\mu\nu}|} - \lambda\sqrt{-g}\right)
     + S_{\rm M}[g,\Psi_{\rm M}],
\end{equation}
where $R_{\mu\nu}$ is the symmetric part of the Ricci tensor constructed with the connection $\Gamma^\mu_{\alpha\beta}$, while $S_{\rm M}$ denotes the matter action depending on the metric and matter field. 
$|g_{\mu\nu} + \kappa R_{\mu\nu}|$ means the absolute value of the determinant of the matrix of $(g_{\mu\nu} + \kappa R_{\mu\nu})$.
It should be remarked that this action for $S_{\rm M}=0$ can recover the Einstein-Hilbert action, i.e., EiBI in vacuum is identical to general relativity \cite{EiBI}. The dimensionless constant $\lambda$ is associated with the cosmological constant as $\Lambda=(\lambda - 1)/\kappa$. In this paper we adopt $\lambda=1$ to focus on the relativistic stars with asymptotically flatness. Additionally, the constraints on the Eddington parameter $\kappa$ are also discussed in term of the solar observations, big bang nucleosynthesis, and the existence of neutron stars \cite{EiBI,kappa01,kappa02,PCD2011}. We should remark that the stellar structures could depend on the value of $\lambda$, which should be considered somewhere.

As the feature of this theory, the metric $g_{\mu\nu}$ and the connection $\Gamma^{\alpha}_{\mu\nu}$ are considered as the independent fields. Then, the field equations can be obtained by varying the action \cite{EiBI};
\begin{gather}
  \Gamma^\mu_{\alpha\beta} = \frac{1}{2}q^{\mu\sigma}\left(q_{\sigma\alpha,\beta} + q_{\sigma\beta,\alpha}
     - q_{\alpha\beta,\sigma}\right), \label{eq:1} \\
  q_{\mu\nu} = g_{\mu\nu} + \kappa R_{\mu\nu}, \label{eq:2} \\
  \sqrt{-q}q^{\mu\nu} = \sqrt{-g}g^{\mu\nu} - 8\pi\kappa\sqrt{-g}T^{\mu\nu}, \label{eq:3}
\end{gather}
where $q_{\mu\nu}$ is an auxiliary metric and $T^{\mu\nu}$ is the standard energy-momentum tensor with indices raised with the metric $g_{\mu\nu}$. In addition to the above field equations, the energy-momentum conservation should be satisfied, i.e., $\nabla_{\mu}T^{\mu\nu}=0$, where $\nabla_\mu$ is defined with the physical metric $g_{\mu\nu}$.
It is noticed that $q^{\mu\nu}$ is the matrix inverse of $q_{\mu\nu}$, which is different from $g^{\mu\alpha}g^{\nu\beta}q_{\alpha\beta}$ if matter exists.

Now, we consider the spherically symmetric relativistic stars. Previously, the structures of compact objects in EiBI have already been examined by several groups \cite{PCD2011,PDC2012,SLL2012,SLL2013,HLMS2013}. The metric describing the spherically symmetric objects can be written as 
\begin{gather}
  g_{\mu\nu}dx^\mu dx^\nu = -e^{\nu(r)}dt^2 + e^{\lambda(r)}dr^2 + f(r)d\Omega^2, \\
  q_{\mu\nu}dx^\mu dx^\nu = -e^{\beta(r)}dt^2 + e^{\alpha(r)}dr^2 + r^2d\Omega^2,
\end{gather}
where $d\Omega^2=d\theta^2 + \sin^2\theta d\phi^2$. We remark that we use the gauge freedom to fix that $q_{\theta\theta}=r^2$. In particular, we consider the neutron stars composed of the perfect fluid, which is given by
\begin{equation}
  T^{\mu\nu} = \left(\epsilon + p\right)u^\mu u^\nu + pg^{\mu\nu},
\end{equation}
where $\epsilon$ and $p$ denote the energy density and pressure, while $u^{\mu}$ corresponds to the four velocity of matter given as $u^{\mu}=(e^{-\nu/2},0,0,0)$. Using Eq. (\ref{eq:3}), one can obtain the relation
\begin{equation}
 abf = r^2, \ \ e^{\alpha} = e^{\lambda}ab, \ \ e^{\beta}=e^{\nu}b^3/a, \label{eq:abf}
\end{equation}
where $a\equiv \sqrt{1+8\pi\kappa \epsilon}$ and $b\equiv \sqrt{1-8\pi\kappa p}$. On the other hand, using Eq. (\ref{eq:2}), one can get the equations describing the structures of relativistic stars;
\begin{gather}
 \left(re^{-\alpha}\right)' = 1-\frac{r^2}{2\kappa}\left[\frac{a}{b^3} - \frac{3}{ab} + 2\right], \\
 e^{-\alpha}\left(1+r\beta'\right) = 1 + \frac{r^2}{2\kappa} \left[\frac{a}{b^3} + \frac{1}{ab} - 2\right],
\end{gather}
where the prime denotes a derivative with respect to $r$. In addition to these equations, the energy-momentum conservation law gives us the additional equation;
\begin{equation}
 \nu' = -\frac{2p'}{\epsilon+p}. \label{eq:nu'}
\end{equation}

At last, combining Eqs. (\ref{eq:abf}) -- (\ref{eq:nu'}), one can derive the Tolman-Oppenheimer-Volkoff (TOV) equations in EiBI;
\begin{gather}
 m' = \frac{r^2}{4\kappa}\left[\frac{a}{b^3} - \frac{3}{ab} + 2\right], \ \  e^{-\alpha} = 1-\frac{2m}{r}, \\
 p' = -e^{\alpha}\left[\frac{2m}{r^2} + \frac{r}{2\kappa}\left(\frac{a}{b^3}+\frac{1}{ab}-2\right)\right] \nonumber \\
    \times \left[\frac{2}{\epsilon+p}+4\pi\kappa\left(\frac{3}{b^2}+\frac{1}{a^2 c_s^2}\right)\right]^{-1},
\end{gather}
where $c_s$ denotes the sound speed. We remark that these equations in the limit of $\kappa\to 0$ can reduce to the standard TOV equations in general relativity. With the relation between $\epsilon$ and $p$, i.e., EOS, the equation system is closed. After one adopts the central density $\epsilon_c$, the TOV equations are integrated outward with the conditions $m(0)=0$. Then, the stellar surface should become the position where the pressure vanishes. Since EiBI is equivalent to general relativity in vacuum and $\epsilon=p=0$ at the stellar surface ($r=R$), one can find that $e^{-\alpha}=e^{-\lambda}=1-2M/R$ at $r=R$. As a result, the stellar mass is defined as $M=m(R)$. Additionally, in order to allow for self-gravitating objects, the condition of $\kappa$ is obtained \cite{PCD2011} as
\begin{gather}
  8\pi p_c\kappa < 1\ \  {\rm for}\ \  \kappa>0, \\
  8\pi\epsilon_c |\kappa| <1\ \  {\rm for} \ \ \kappa<0,
\end{gather}
where $p_c$ denotes the central pressure. Hereafter, we adopt $8\pi\epsilon_0\kappa$ as a normalized constant, where $\epsilon_0$ is the nuclear saturation density given by $2.68\times 10^{14}$ g cm$^{-3}$. We remark that $\epsilon_0=1.99\times 10^{-4}$ km$^{-2}$ in geometric units with $c=G=1$.

\begin{figure}[t]
\begin{center}
\includegraphics[scale=0.53]{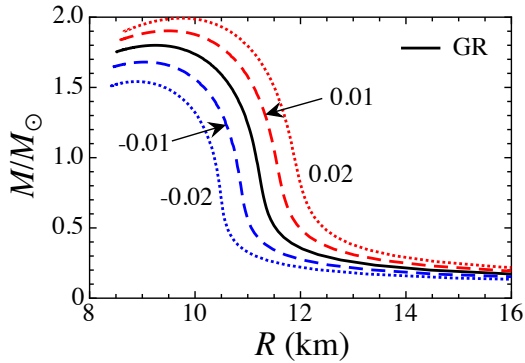} 
\end{center}
\caption{
(Color online) Neutron star mass-radius relations in EiBI constructed from FPS EOS. The labels on lines denote the values of $8\pi\epsilon_0\kappa$. The solid line corresponds to that in general relativity.
}
\label{fig:MR}
\end{figure}

\section{Relativistic stellar models in EiBI}
\label{sec:III}
In order to construct relativistic stellar models, we need to prepare EOS. In this paper, we adopt the realistic EOSs proposed by the different theoretical approaches, i.e., the phenomenological models, the relativistic mean field models, and the ones based on the Skyrme-type effective interactions (see \cite{SIOO2014} for more information about EOSs adopted here). As the phenomenological models, we adopt the EOS constructed by Oyamatsu and Iida \cite{OI2003,OI2007}, where they made EOS for various values of incompressibility $K_0$ and the density dependence of the nuclear symmetry energy at the saturation point $L$. $K_0$ and $L$ are parameters characterizing the stiffness of neutron-rich nuclear matter. Hereafter, we refer to this phenomenological EOS as OI ($a,b$), where $a$ and $b$ denote the adopted values of $K_0$ and $L$. As the relativistic mean field models, we adopt two EOSs, i.e., Shen EOS \cite{ShenEOS} and Miyatsu EOS \cite{Miyatsu}. We also adopt five EOSs based on the Skyrme-type effective interactions, i.e., FPS \cite{FPS}, SLy4 \cite{SLy4}, BSk19, BSk20, and BSk21 \cite{BSk,PGC2011,PCGD2012}. We remark that every EOS adopted in this paper is consistent with the terrestrial experimental data for masses and radii of stable nuclei. This is important point to consider the neutron stars with $0.5M_\odot$, because the density inside such objects is less than a few times saturation density, which should be strongly constrained from the terrestrial experiments \cite{SIOO2014}.

As a example of neutron star models in EiBI, we show the mass and radial relations constructed from FPS EOS in Fig. \ref{fig:MR}, where the solid line denotes the results in general relativity ($\kappa =0$), while the broken and dotted lines correspond to those in EiBI with $8\pi\epsilon |\kappa| = 0.01$ and 0.02, respectively. From this figure, one can observe the obvious deviation from the predictions in general relativity. However, as mentioned the above, this difference depending on the coupling constant $\kappa$ must be buried in the uncertainties due to EOS of neutron star matter. That is, it could be quite difficult to distinguish EiBI from general relativity only if one would measure the mass and radius of neutron stars.

\begin{figure}[t]
\begin{center}
\includegraphics[scale=0.53]{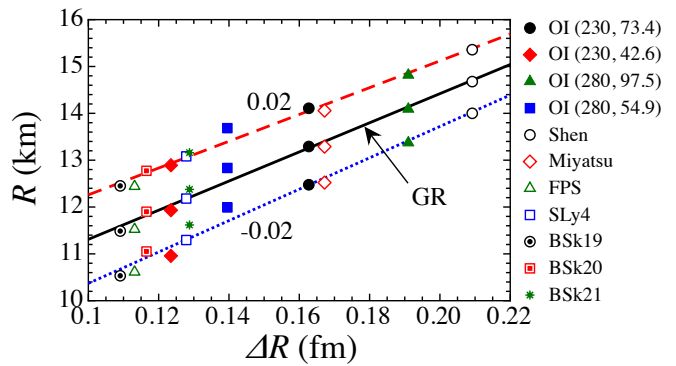} 
\end{center}
\caption{
(Color online) Radii of neutron stars with $0.5 M_\odot$, $R_{05}$, as a function of neutron skin thickness of ${}^{208}$Pb for $8\pi\epsilon_0\kappa=-0.02$, 0, and 0.02, using the various EOSs. The solid line denotes the fitting line in general relativity, while the broken and dotted lines denote that in EiBI for $8\pi\epsilon_0\kappa=0.02$ and $-0.02$, respectively.
}
\label{fig:skin-R}
\end{figure}

With respect to such a difficulty, we are successful to find observational possibility to discriminate EiBI from general relativity, i.e., via the terrestrial experiments for the neutron skin thickness of neutron-rich atomic nuclei. Using the various realistic EOSs mentioned the above, we determine the radii of neutron stars with $0.5M_\odot$, as varying the value of $8\pi\epsilon_0\kappa$, and show it in Fig. \ref{fig:skin-R} as a function of neutron skin thickness of ${}^{208}$Pb, where $R_{05}$ and $\Delta R$ denote the stellar radii with $0.5M_\odot$ and the neutron skin thickness of ${}^{208}$Pb.
In particular, in order to estimate the value of $\Delta R$ for each EOS, we adopt the formula proposed by Oyamatsu and Iida \cite{OI2003}, where the neutron skin thickness can be expressed as functions of neutron excess, atomic mass number, and value of $L$. Since the estimation of $\Delta R$ could depend a little on theoretical models, the plots in Fig. \ref{fig:skin-R} may be slightly modified. Anyway, the value of $\Delta R$ dose not depend on $\kappa$ at all.
From this figure, one clearly observes that $R_{05}$ can be written as a linear function of $\Delta R$ almost independently of the adopted EOS, while the correlation between $R_{05}$ and $\Delta R$ strongly depends on the coupling constant $\kappa$. In practice, one can write the linear fitting with each value of $8\pi\epsilon_0\kappa$ as
\begin{equation}
  R_{05} = c_0 + c_1 \Delta R, \label{eq:fit}
\end{equation}
where $c_0$ and $c_1$ are constants depending on the value of $8\pi\epsilon_0\kappa$. Since the units of $R_{05}$ and $\Delta R$ in this fitting are km and fm, the units of $c_0$ and $c_1$ become km and km/fm. Regarding the suitability of this fitting, we can estimate that the root mean fractional variation (RMFV) from the original values are $1.44$ \%, $1.09$ \%, and $0.90$ \% for $8\pi\epsilon_0\kappa=-0.02$, 0, and 0.02, respectively.
Here, RMFV for each $\kappa$ is calculated as
\begin{equation}
  {\rm RMFV} = \sqrt{\frac{1}{N}\sum_N\frac{\left(R^{\rm ex}_{05} - R_{05}^{\rm (\ref{eq:fit})}\right)^2}{{R^{\rm ex}_{05}}^2}},
\end{equation}
where $R_{05}^{\rm ex}$ and $R_{05}^{\rm (\ref{eq:fit})}$ denote the stellar radius with $0.5M_\odot$ calculated with each EOS and that estimated with Eq. (\ref{eq:fit}), while $N$ is the number of adopted EOSs, i.e., $N=11$ in this paper. The number of RMFV expresses how good the fits are. 
Thus, we consider that the linear fitting (\ref{eq:fit}) is enough to distinguish the gravitational theory.

\begin{figure}[t]
\begin{center}
\includegraphics[scale=0.53]{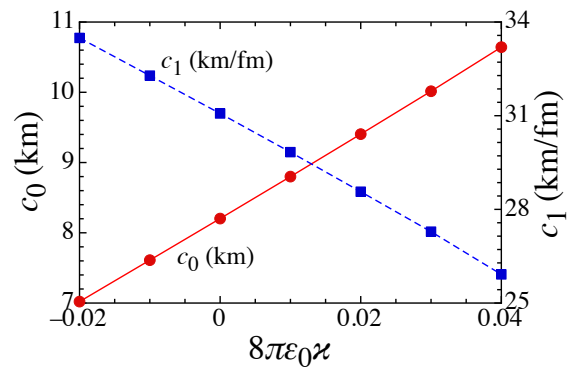} 
\end{center}
\caption{
(Color online) Coefficients in the fitting formula (\ref{eq:fit}) as a function of $8\pi\epsilon_0\kappa$, where the circles and squares denote the obtained coefficients $c_0$ and $c_1$, respectively.
}
\label{fig:c0c1}
\end{figure}

Moreover, we also examine the dependence of $c_0$ and $c_1$ on $8\pi\epsilon_0\kappa$ in the range of $-0.02\le 8\pi\epsilon_0\kappa\le 0.04$. Fig. \ref{fig:c0c1} shows the values of $c_0$ and $c_1$ as a function of $8\pi\epsilon_0\kappa$. From this figure, one finds that the coefficients in the linear fitting (\ref{eq:fit}), i.e., $c_0$ and $c_1$, can be written as linear functions of $8\pi\epsilon_0\kappa$. In fact, we can derive such linear functions as
\begin{gather}
  c_0/{\rm km} = 8.21 + 60.3\times(8\pi\epsilon_0\kappa), \label{eq:c0} \\
  c_1/{\rm (km/fm)} = 31.0 - 125.8\times(8\pi \epsilon_0\kappa). \label{eq:c1}
\end{gather}
Consequently, combining Eqs. (\ref{eq:fit}), (\ref{eq:c0}), and (\ref{eq:c1}), one can obtain the value of $8\pi\epsilon_0\kappa$ as a function of $R_{05}$ and $\Delta R$;
\begin{equation}
  8\pi\epsilon_0\kappa = \frac{(R_{05}/{\rm km}) - 8.21 - 31.0(\Delta R/{\rm fm})}{60.3 - 125.8(\Delta R/{\rm fm})}. \label{eq:kappa}
\end{equation}
Using this empirical formula, at least, one must be able to check how valid general relativity is. Namely, with the observational values of $R_{05}$ and $\Delta R$, one can estimate the value of $\kappa$, where $\kappa$ should be zero in general relativity. On the other hand, we should also emphasize that this empirical formula could be adopted to distinguish EiBI from most of the modified theories of gravity, because the stellar properties with $0.5M_\odot$ in most of the modified theories of gravity are almost same as those in general relativity.

At last, we should also mention an uncertainty in $8\pi\epsilon_0\kappa$ due to the observational uncertainties in $R_{05}$ and $\Delta R$. If the observations of $R_{05}$ and $\Delta R$ have $\pm 10\%$ variances, one can estimate that the variances on $8\pi\epsilon_0\kappa$ arise up to $\pm 0.04$ for $R_{05}=12$ km and $\pm 0.06$ for $R_{05}=14$ km, using Eq. (\ref{eq:kappa}), even if general relativity is correct gravitational theory.



\section{Current Observational Status}
\label{sec:IV}

In order to estimate the coupling constant $\kappa$ with Eq. (\ref{eq:kappa}), one needs to measure the neutron skin thickness of ${}^{208}$Pb, $\Delta R$, and the radius of $0.5M_\odot$ neutron star, $R_{05}$. That is, depending on some finite precision in the measurements of $\Delta R$ and $R_{05}$, the estimation of $\kappa$ could become blurry. In this section, we discuss the current observational status and how well $\kappa$ may be constrained with $\Delta R$ and $R_{05}$.

The current best measurement of $\Delta R$ could be the data by PREX experiment \cite{deltaR}, which tells us that $\Delta R=0.33^{+0.16}_{-0.18}$ fm. Unfortunately, within this precision of $\Delta R$, we can not constrain $\kappa$ even if the exact value of $R$ is known. In fact, since $\Delta R=0.33$ fm is out of Fig. \ref{fig:skin-R}, it may be questionable to adopt Eq. (\ref{eq:kappa}) in such region to estimate $\kappa$. 

The measurement of $R_{05}$ must be more difficult than that of $\Delta R$, because $0.5M_\odot$ is an exceptionally small mass for neutron star and the measurement of stellar radius itself is quite challenging. The masses of neutron stars in binary system have been determined \cite{OPNV2012,KKYT2014}. Among them, the lowest mass of neutron star observed so far could be $0.87\pm 0.07M_\odot$ with eccentric orbit or $1.00\pm 0.10M_\odot$ with circular orbit observed in the high-mass X-ray binary 4U1538-52 \cite{4u1538}. We hope that the measurement of an extremely low-mass neutron star will be successful in the future. In that time, perhaps, we may not need the exactly $0.5M_\odot$ neutron star, because the stellar radius is theoretically predicted to be quasi-constant for the neutron star with $M=0.5-0.7M_\odot$, where the difference in radius could be less than a few \% \cite{LP2001}. 
In fact, in a similar way in section \ref{sec:III}, we find that the radii of $0.7M_\odot$ stellar models, $R_{07}$, can be also expressed as a linear function of $\Delta R$ almost independently of the adopted EOSs, although the values of RMFV become slightly worse, i.e. 2.00\%, 1.44\%, and 1.08\% for $8\pi\epsilon_0\kappa=-0.02$, 0, and 0.02, respectively. Using the obtained linear relation between $R_{07}$ and $\Delta R$ with Eq. (\ref{eq:fit}), the relative deviations between $R_{05}$ and $R_{07}$ for $\Delta R=0.1$ fm become 1.16\%, 2.24\%, and 3.32\%, while those for $\Delta R=0.22$ fm become  0.33\%, 0.79\%, and 1.41\% for $8\pi\epsilon_0\kappa=-0.02$, 0, and 0.02, respectively.  Thus, one might estimate the value of $8\pi\epsilon_0\kappa$ on some level from Eq. (\ref{eq:kappa}) with $R_{07}$ instead of $R_{05}$.


On the other hand, the measurement of stellar radius is notoriously difficult. Recently, the stellar radius of neutron star have been estimated with the observations of thermonuclear X-ray bursts and thermal spectra from quiescent low-mass X-ray binaries \cite{GSWR2013,LS2013,OBG2010,SLB2012}. However, it is still quite difficult to determine the stellar radius with less than $\sim\pm 10\%$ accuracy. 
In order to determine the value of $8\pi\epsilon_0\kappa$ within the range of $\pm 0.01$, one should measure the stellar radius at least with $\pm$ a few \% accuracy, even if $\Delta R$ will be exactly measured.

\section{Conclusion}
\label{sec:V}
Compact objects must be suitable laboratories to test the gravitational theory. However, due to the uncertainties of EOS of neutron star matter, it is generally difficult to distinguish the gravitational theory by only using the observations of neutron stars. To solve this difficulty, we find observational possibility to discriminate EiBI from general relativity via the measurements of neutron skin thickness of ${}^{208}$Pb by the terrestrial experiments, in addition to the astronomical observations of neutron stars. We show that the stellar radii with $0.5M_\odot$ can be written as a linear function of neutron skin thickness almost independently of the adopted EOS, which strongly depends on the coupling constant in EiBI. Additionally, we show the coupling constant can be estimated with the observable properties, such as the stellar radii with $0.5M_\odot$ and neutron skin thickness of ${}^{208}$Pb. This estimation could be also useful to observationally probe how valid general relativity is.

In this paper, we especially focus on the neutron skin thickness of ${}^{208}$Pb, but the linear correlation between the stellar radii with $0.5M_\odot$ and neutron skin thickness could be satisfied for the different neutron-rich nuclei. If so, one could make estimations of the coupling constant in EiBI via the different measurements of neutron skin thickness of various nuclei, which tells us more accurate value of the coupling constant. On the other hand, it could be difficult to discuss the similar correlation by using the radius with massive neutron star, because the density inside the massive neutron star becomes much higher than the nuclear saturation density, where there exist many theoretical uncertainties to construct EOS models and the additional components such as hyperon and/or quark could appear.


\acknowledgments
We are grateful to K. Yagi, K. Iida, K. Oyamatsu, and our referees for their fruitful comments. We also acknowledge the hospitality of members of University of Mississippi, especially E. Berti. This work was supported
by Grants-in-Aid for Scientific Research on Innovative Areas through No.\ 24105001 and No.\ 24105008 provided by MEXT, 
by Grant-in-Aid for Young Scientists (B) through No.\ 24740177 and No.\ 26800133 provided by JSPS,
by the Yukawa International Program for Quark-hadron Sciences,
by the Grant-in-Aid for the global COE program ``The Next Generation of Physics, Spun from Universality and Emergence" from MEXT,
and
by a grant from Yukawa Institute for Theoretical Physics, Kyoto University.



\end{document}